\documentclass[aps,preprint,nofootinbib]{revtex4-2}

\usepackage[top=2.5 cm, bottom=2.5 cm, left=2.5 cm, right=2.5 cm]{geometry}

\usepackage[utf8]{inputenc}
\usepackage{amsfonts}
\usepackage{color}

\usepackage{amssymb}
\usepackage{amsmath}
\usepackage{stmaryrd}
\usepackage{latexsym}
\usepackage{tcolorbox}
\usepackage{multirow}

\usepackage{caption}
\usepackage{subcaption}

\usepackage{xcolor}
\definecolor{myurlcolor}{HTML}{08457E}
\definecolor{mylinkcolor}{HTML}{2A52BE}
\definecolor{mycitecolor}{HTML}{E30022}

\usepackage{float}
\usepackage[colorlinks, linkcolor=mylinkcolor, citecolor=mycitecolor, urlcolor=myurlcolor, linktocpage=true]{hyperref}

\def\equationautorefname~#1\null{(#1)\null}
\def\tableautorefname~#1\null{(#1)\null}
\def\figureautorefname~#1\null{(#1)\null}
\def\sectionautorefname~#1\null{(#1)\null}

\let\origref\autoref
\def\autoref#1{\textbf{\origref{#1}}}

\let\origcite\cite
\def\cite#1{\textbf{\origcite{#1}}}

\usepackage{tikz}
\usepackage{pgfplots}
\pgfplotsset{compat=1.15}

\usepackage{titlesec}
\titleformat*{\section}{\centering\small\bfseries\scshape}
\titleformat*{\subsection}{\small\bfseries\scshape}
\titleformat*{\subsubsection}{\small\bfseries\scshape}

\usepackage{array}
\newcommand{\PreserveBackslash}[1]{\let\temp=\\#1\let\\=\temp}
\newcolumntype{C}[1]{>{\PreserveBackslash\centering}p{#1}}
\newcolumntype{R}[1]{>{\PreserveBackslash\raggedleft}p{#1}}
\newcolumntype{L}[1]{>{\PreserveBackslash\raggedright}p{#1}}

\def\H{{\cal H}}      % script H for Hamiltonian
\def\R{{\cal R}}     % script R for
\def\V{{\cal V}}     % script R for
\def\T{{\cal T}}     % script R for
\def\P{{\cal P}}     % script R for
\renewcommand{\d}{\mathrm{d}}
\newcommand{\Mpl}{M_{\mathrm{Pl}}}
\newcommand{\mpl}{m_{\mathrm{Pl}}}

\newcommand{\sfrac}[2]{\dfrac{\,#1\,}{\,#2\,}}

\newcommand{\der}[2]{\sfrac{d #1}{d #2}}
\newcommand{\derp}[2]{\sfrac{\partial #1}{\partial #2}}

\newcommand{\mf}[1]{\mathbf{#1}}
\newcommand{\nb}{\nabla}
\newcommand{\sg}{\sigma}
\newcommand{\dt}{\delta}

\newcommand{\lb}{\lambda}
\newcommand{\al}{\alpha}
\newcommand{\bt}{\beta}

\newcommand{\Om}{\Omega}

\newcommand{\p}{\partial}

\let\oldsqrt\sqrt
\def\sqrt{\mathpalette\DHLhksqrt}
\def\DHLhksqrt#1#2{%
	\setbox0=\hbox{$#1\oldsqrt{#2\,}$}\dimen0=\ht0
	\advance\dimen0-0.4\ht0
	\setbox2=\hbox{\vrule height\ht0 depth -\dimen0}%
	{\box0\lower0.4pt\box2}}

\usepackage[verbose]{placeins}

\begin{document}

\title{Cosmological First-Order Vacuum Phase Transitions \\ in an Expanding Anisotropic Universe \vspace{5mm}}

\author{A.\ Savaş Arapoğlu}
\email{arapoglu@itu.edu.tr}

\author{A.\ Emrah Yükselci}
\email{yukselcia@itu.edu.tr}

\affiliation{\vspace{5mm}Istanbul Technical University, Faculty of Science and Letters, Physics Engineering Department, 34469, Maslak, Istanbul, Turkey \vspace{2cm}}

\begin{abstract}\vspace{5mm}
We examine the anisotropy originated from a first-order vacuum phase transitions through three-dimensional numerical simulations. We apply Bianchi Type-I metric to our model that has one scalar field minimally coupled to the gravity. We calculate the time evolution of the energy density for the shear scalar and the directional Hubble parameters as well as the power spectra for the scalar field and the gravitational radiation although there are a number of caveats for the tensor perturbations in Bianchi Type-I universe. We run simulations with different mass scales of the scalar field, therefore, in addition to investigation of anisotropy via the shear scalar, we also determine at which mass scale the phase transition completes successfully, hence, neglecting the expansion of the Universe does not significantly affect the results. Finally, we showed that such an event may contribute to the total anisotropy depending on the mass scale of the scalar field and the initial population of nucleated bubbles.
\end{abstract}

\maketitle
\newpage
\raggedbottom
%\tableofcontents

\section{INTRODUCTION\label{sec:intro}}

The first direct detection of the gravitational waves (GWs) \cite{First_GW_detection} has opened a new era in observation of the Universe since they can carry the information related to very source phenomenon thanks to their weakly-interacting character. Although recent results of the observation with low-frequency GWs may point out an astrophysical origin \cite{NANOGrav2023} as is the first one, yet this is another important step towards mapping the stochastic GW background that may have contributions originated from the events of the early stages in the Universe as well. The imprints of such GWs may be detected through space-based GW detectors that are planned to be built in the future \cite{LISA,LISA_rev,DECIGO,TAIJI,TIANQIN,BBO}.

Cosmological first-order phase transitions (PTs), which may have possibly occurred in the early Universe, are one type of phenomenon that can create GWs as an outcome \cite{Mazumdar2019,Hindmarsh2020}. In spite of the fact that the well-known examples are the electroweak and the quark-hadron PTs, they may have taken place at any scale in the early Universe between QCD \cite{QCD_scale} and GUT \cite{GUT_scale} scales, respectively. However, the standard model of particle physics does not predict first-order phase transitions \cite{Kajantie1996_1,Kajantie1996_2}, yet there are many extensions of it to allow that (see e.g. \cite{Mazumdar2019,Hindmarsh2020,Weir2017} and references therein). This type of event could take place through the bubble nucleation mechanism, theory of which was studied at zero \cite{Coleman1977_1, Coleman1977_2} and finite \cite{Linde1980, Linde1983} temperatures in flat space-time whereas the gravitational effects were discussed later in Ref. \cite{Coleman1980}. It is known that the most probable initial profile for the bubbles has a $O(4)$ symmetric form \cite{Coleman1978} although there is no such proof for a curved space-time yet.

The GWs originated from first-order PTs are created by the shear stress caused by the deformation of the symmetric structure of colliding bubbles. The theoretical approach to examine such an event was studied in Ref.\ \cite{Turner1992_env} where collided portions of the bubbles are not taken into account. However, it was shown through the numerical simulations \cite{Cutting2018,Cutting2019,Cutting2020} that those parts should be considered since the scalar field oscillates around its true vacuum and give rise to another peak in the gravitational radiation power spectrum related to its mass in addition to the maximum value associated with the mean separation between bubbles. This has the potential to determine the parameters of a model and/or even the model itself. Recently, in the context of the scalar-tensor theories, the scalar field non-minimally coupled to the gravity has also been studied through the numerical simulations \cite{Arapoglu2022}. Another example is the study of the two-step phase transition related to electroweak symmetry breaking \cite{Zhao2022}. In addition to numerical approaches, it has been shown that it is also possible to analytically calculate the power spectra to some extent for the gravitational radiation formed during the bubble collision phase \cite{Caprini2007,Jinno2017,Zhong2021}.

In order to investigate the anisotropy we implement Bianchi Type-I metric where each direction has a different scale factors unlike the Friedmann–Lemaître–Robertson–Walker metric which is, indeed, a particular type of Bianchi Type-I model in this manner. This model has been widely used in the cosmological context (see e.g. \cite{Clifton2011} and references therein) since it is one of the simplest extensions for isotropic space-time and it may even offer some solutions to the well-known problems such as $H_0$ tension \cite{Akarsu2019}. Moreover, on the GW front with another aspect of the anisotropy, a possible detection of it in the stochastic GW background may even enable to distinguish between superimposed sources \cite{LISA2022}. On the other hand, in light of recent observations, it has been reported that any sort of anisotropy is not encountered in the data \cite{NANOGrav2023_anisotropy}. However, the picture may change after the inclusion of more data into the analysis, and this may lead to understand the birthplace of an observed GW signal, in other words, whether it has an astrophysical or cosmological origins. Although there is no concrete evidence yet to determine the origins of the signal \cite{Bian2023,Figueroa2023}, analysis in Ref.\ \cite{Wu2023} on the NanoGRAV data \cite{NANOGrav2023} indicates that some cosmological sources, e.g. strong first-order PTs, can provide comparable results with the astrophysical ones such as supermassive black hole binaries.

The paper is structured as follows: In Section \autoref{sec:setup}, we describe the main equations of the model; in Section \autoref{sec:numerical_setup} we provide the modified equations in accordance with the numerical scheme; in Section \autoref{sec:observables} we define the quantities to be followed during the simulations; in Section \autoref{sec:results} we present the outcomes of the simulations and discuss the results in Section \autoref{sec:conclusion}.

\section{SET-UP\label{sec:setup}}

In this section we provide the main equations that will be used throughout the paper. To this end, we start with the Einstein field equations given as
\begin{equation}
    \R_{\mu\nu} - \sfrac{1}{2} \R g_{\mu\nu} = \Mpl^{-2} \, \T_{\mu\nu}
\label{eq:field_eq_einstein}
\end{equation}
where the Planck mass is defined via $\Mpl^{-2} \equiv 8\pi G$ and is equal to $\Mpl = 2.435 \times 10^{18}$ GeV in natural units, i.e. $c=\hbar=1$, which is adopted in this work. In the presence of only one scalar field minimally coupled to the gravity the energy-momentum tensor is given by the following expression
\begin{equation}
    \T_{\mu\nu} = \nb_{\!\mu} \, \phi \, \nb_{\!\nu} \, \phi - \sfrac{1}{2} \, g_{\mu\nu} \, \nb^\sg \phi \, \nb_{\!\!\sg} \, \phi - g_{\mu\nu}\,V(\phi) \:.
\label{eq:en_mom_tensor}
\end{equation}
On the other hand, the equation of motion for the scalar field is obtained as
\begin{align}
    \nb^\sg \nb_{\!\!\sg} \, \phi - \derp{V\!(\phi)}{\phi} = 0
\label{eq:field_eq_scalar}
\end{align}
and we use the potential in the form of 
\begin{equation}
    V(\phi) = \sfrac{1}{2} M^2 \phi^2 + \sfrac{1}{3} \dt \phi^3 + \sfrac{1}{4} \lb \phi^4 - V_{\rm{c}}
\label{eq:potential}
\end{equation}
where $M$, $\dt$, $\lb$, and $V_{\rm{c}}$ are constants.

For the background evolution, we implement the Bianchi Type-I metric as
\begin{equation}
    \d s^2 = -\d t^2 + a_1^2(t) \, \d x^2 + a_2^2(t) \, \d y^2 + a_3^2(t) \, \d z^2 
\label{eq:metric}
\end{equation}
where $a_1(t)$, $a_2(t)$, and $a_3(t)$ are the scale factors in $x$, $y$, and $z$ directions and they are the functions of time only. For future convenience, we also define 
\begin{align}
    H_i \equiv \sfrac{\dot{a}_i}{a_i} \:\:\: &(i=1,2,3) \:, \qquad\quad H \equiv \sfrac{1}{3} \big( H_1 + H_2 + H_3 \big)
\label{eq:hubble_parameters}
\end{align}
where $H_i$ is the directional and $H$ is the average Hubble parameters.

The metric given in Eq.\ \autoref{eq:metric} yields $tt$, $xx$, $yy$, and $zz$ components of Eq.\ \autoref{eq:field_eq_einstein}, respectively, in the following forms
\begin{align}
    \sfrac{\dot{a}_1}{a_1} \sfrac{\dot{a}_2}{a_2} + \sfrac{\dot{a}_1}{a_1} \sfrac{\dot{a}_3}{a_3} + \sfrac{\dot{a}_2}{a_2} \sfrac{\dot{a}_3}{a_3} &= \Mpl^{-2} \left[ \sfrac{1}{2} \left( \langle\dot{\phi}^2\rangle + \sfrac{\langle\phi_x^2\rangle}{a_1^2} + \sfrac{\langle\phi_y^2\rangle}{a_2^2} + \sfrac{\langle\phi_z^2\rangle}{a_3^2} \right) + \langle V(\phi)\rangle \right] \label{eq:eins00} \\[2mm]
    \sfrac{\Ddot{a}_2}{a_2} + \sfrac{\Ddot{a}_3}{a_3} + \sfrac{\dot{a}_2}{a_2} \sfrac{\dot{a}_3}{a_3} &= \Mpl^{-2} \left[ \sfrac{1}{2} \left( - \langle\dot{\phi}^2\rangle - \sfrac{\langle\phi_x^2\rangle}{a_1^2} + \sfrac{\langle\phi_y^2\rangle}{a_2^2} + \sfrac{\langle\phi_z^2\rangle}{a_3^2} \right) + \langle V(\phi)\rangle \right] \label{eq:eins11} \\[2mm]
    \sfrac{\Ddot{a}_1}{a_1} + \sfrac{\Ddot{a}_3}{a_3} + \sfrac{\dot{a}_1}{a_1} \sfrac{\dot{a}_3}{a_3} &= \Mpl^{-2} \left[ \sfrac{1}{2} \left( - \langle\dot{\phi}^2\rangle + \sfrac{\langle\phi_x^2\rangle}{a_1^2} - \sfrac{\langle\phi_y^2\rangle}{a_2^2} + \sfrac{\langle\phi_z^2\rangle}{a_3^2} \right) + \langle V(\phi)\rangle \right] \label{eq:eins22} \\[2mm]
    \sfrac{\Ddot{a}_1}{a_1} + \sfrac{\Ddot{a}_2}{a_2} + \sfrac{\dot{a}_1}{a_1} \sfrac{\dot{a}_2}{a_2} &= \Mpl^{-2} \left[ \sfrac{1}{2} \left( - \langle\dot{\phi}^2\rangle + \sfrac{\langle\phi_x^2\rangle}{a_1^2} + \sfrac{\langle\phi_y^2\rangle}{a_2^2} - \sfrac{\langle\phi_z^2\rangle}{a_3^2} \right) + \langle V(\phi)\rangle \right] \label{eq:eins33}
\end{align}
where the angle brackets denotes the spatial average over all simulation box, the dot represents the derivative with respect to $t$, and the subscript letters $x,y,z$ stand for the spatial derivatives in the corresponding directions. Moreover, for the sake of simplification we eliminate the time derivative of the scalar field from Eqs.\ \autoref{eq:eins11}, \autoref{eq:eins22}, \autoref{eq:eins33} with the help of the constraint equation, i.e. Eq.\ \autoref{eq:eins00}, and obtain the final form of the equations of motion for the scale factors as follows
\begin{align}
    \sfrac{\Ddot{a}_1}{a_1} + \sfrac{\dot{a}_1}{a_1} \bigg( \sfrac{\dot{a}_2}{a_2} + \sfrac{\dot{a}_3}{a_3} \bigg) &= \Mpl^{-2} \left[ \sfrac{\langle\phi_x^2\rangle}{a_1^2} + \langle V(\phi)\rangle \right] \\[2mm]
    \sfrac{\Ddot{a}_2}{a_2} + \sfrac{\dot{a}_2}{a_2} \bigg( \sfrac{\dot{a}_1}{a_1} + \sfrac{\dot{a}_3}{a_3} \bigg) &= \Mpl^{-2} \left[ \sfrac{\langle\phi_y^2\rangle}{a_2^2} + \langle V(\phi)\rangle \right] \\[2mm]
    \sfrac{\Ddot{a}_3}{a_3} + \sfrac{\dot{a}_3}{a_3} \bigg( \sfrac{\dot{a}_1}{a_1} + \sfrac{\dot{a}_2}{a_2} \bigg) &= \Mpl^{-2} \left[ \sfrac{\langle\phi_z^2\rangle}{a_3^2} + \langle V(\phi)\rangle \right] \:.
\end{align}

On the other hand, the equation of motion for the scalar field, Eq.\ \autoref{eq:field_eq_scalar}, becomes
\begin{align}
    \ddot{\phi} + 3H \dot{\phi} - \sum^3_{k=1} \sfrac{\p^2_k \phi}{a_k^2} + \derp{V\!(\phi)}{\phi} = 0
\end{align}
where the average Hubble parameter, $H$, is defined in Eq.\ \autoref{eq:hubble_parameters} and the potential for the scalar field is given in Eq.\ \autoref{eq:potential}.

Regarding the gravitational waves, the transverse-traceless (TT) part of the tensor perturbations, $h_{ij}$, can be related to an auxiliary tensor, $u_{ij}$, by defining a projection operator \cite{Figueroa2007} as 
\begin{equation}
    h_{ij}(t,\mf{k}) = \Lambda_{ij,lm}(\mf{\hat{k}}) \, u_{lm}(t,\mf{k})
\end{equation}
where $u_{lm}(t,\mf{k})$ is the Fourier transform of $u_{ij}(t,\mf{x})$ and the projection operator is defined as
\begin{equation}
    \Lambda_{ij,lm}(\mf{\hat{k}}) = P_{im}(\mf{\hat{k}}) P_{jl}(\mf{\hat{k}}) - \sfrac{1}{2} P_{ij}(\mf{\hat{k}}) P_{lm}(\mf{\hat{k}}) \:, \qquad P_{ij}(\mf{\hat{k}}) = \dt_{ij} - \sfrac{k_i k_j}{k^2} \:.
\end{equation}
Then, this method yields the equation of motion in the following form
\begin{equation}
    \Ddot{u}_{ij} + 3 H \dot{u}_{ij} - \sum^3_{k=1} \sfrac{\p^2_k u_{ij}}{a_k^2} = \Mpl^{-2} \, \sfrac{\p_i \phi \, \p_j \phi}{a_i \, a_j} \:.
\label{eq:aux_tensor}
\end{equation}
Although we use TT gauge for the tensor perturbations here, this not need to be entirely true since there is a gauge fixing problem in Bianchi Type-I model \cite{Miedema1993,Cho1995}. Therefore, we should note that the results related to the tensor perturbations, i.e. the gravitational waves, represented in this work are valid up to a gauge transformation. We will make some more comments on this issue in the following sections.

\vspace{10mm}

\section{NUMERICAL PROCEDURE \label{sec:numerical_setup}}

The code which we used for this work is the same with that of Ref.\ \cite{Arapoglu2022}. It has been written in \texttt{Python} programming language with the help of \texttt{Cython} \cite{Cython} extension for the intensive iterations. Parallel processing has been realized by the pencil decomposition and the communication between processes has been ensured by \texttt{mpi4py} \cite{mpi4py} package. We constructed similar algorithms given in Ref.\ \cite{Mortensen2016} for the Fourier transforms.

We implement the staggered leapfrog algorithm for advance in time with 7-point stencil for the Laplacian operator. In accordance with this scheme, it is necessary to eliminate the first time derivatives of the variables in the equations of motion to achieve the stability and the consistency for the numerical calculations. To this end, we define the new variables and the constants as 
\begin{align}
    &\psi \equiv \sqrt{a_1 a_2 a_3} \, \sfrac{\phi}{\phi_{\rm{t}}} \:, \qquad v_{ij} \equiv \sfrac{\Mpl^2}{\phi_{\rm{t}}^2} \sqrt{a_1 a_2 a_3} \; u_{ij} \:, \qquad U \equiv \sfrac{1}{\phi_{\rm{t}}^2 M^2} (a_1a_2a_3) V \:, \\[2mm]
    &d\tau \equiv M dt \:, \qquad \mf{r} \rightarrow M \mf{r} \:, \qquad \mpl \equiv \Mpl/M \:, \qquad \al \equiv \dt/M \:, \qquad \bt \equiv \phi_{\rm{t}}/M \:.
\end{align}
We will denote the derivative with respect to $\tau$ by the prime symbol in the following sections whereas we keep the same notation for the spatial derivatives, in other words, the subscript $x$ should be understood as the derivative with respect to $Mx$.

In addition to that, we use fixed spatial resolution with $dx=0.44$ in general and the value of the Courant factor is taken $c=0.4$ for all simulations.

\subsection{Numerical Equation Set}

With definition of the new variables the equation of motion for the scalar field given in Eq.\ \autoref{eq:field_eq_scalar} takes the following form
\begin{align}
    \psi'' + K \psi - \sum^3_{k=1} \sfrac{\p^2_k \psi}{a_k^2} + \derp{U\!(\psi)}{\psi} = 0
\label{eq:scalar_field_numerical}
\end{align}
where
\begin{align}
    K \equiv \sfrac{1}{4} \bigg[ \sfrac{a_1^{\prime 2}}{a_1} + \sfrac{a_2^{\prime 2}}{a_2} + \sfrac{a_3^{\prime 2}}{a_3} \bigg] - \sfrac{1}{2} \bigg[ \sfrac{a''_1}{a_1} + \sfrac{a''_2}{a_2} + \sfrac{a''_3}{a_3} + \sfrac{a'_1}{a_1} \sfrac{a'_2}{a_2} + \sfrac{a'_1}{a_1} \sfrac{a'_3}{a_3} + \sfrac{a'_2}{a_2} \sfrac{a'_3}{a_3} \bigg]
\label{eq:coeff_scales}
\end{align}
and the redefined potential is given by
\begin{align}
    U(\psi) = \sfrac{1}{2} \psi^2 + \sfrac{1}{3} \sfrac{\al \bt}{\sqrt{a_1a_2a_3}} \psi^3 + \sfrac{1}{4} \sfrac{\lb \bt^2}{a_1a_2a_3} \psi^4 - \sfrac{a_1a_2a_3}{\bt^2} V_{\rm{c}}
\end{align}
where we set $V_{\rm{c}}$ such that $V(\phi_{\rm{t}})=0$. One may also choose a small constant instead of zero potential value as the cosmological constant. However, this is not in the scope of this paper and it needs to be considered with more realistic setups for long-time simulations.

On the other hand, the equations for the scale factors become
\begin{align}
    \sfrac{a''_1}{a_1} + \sfrac{a'_1}{a_1} \bigg( \sfrac{a'_2}{a_2} + \sfrac{a'_3}{a_3} \bigg) &= \sfrac{\bt^2}{\mpl^2} (a_1a_2a_3)^{-1} \left[ \sfrac{\langle\psi_x^2\rangle}{a_1^2} + \langle U(\psi)\rangle \right] \\[2mm]
    \sfrac{a''_2}{a_2} + \sfrac{a'_2}{a_2} \bigg( \sfrac{a'_1}{a_1} + \sfrac{a'_3}{a_3} \bigg) &= \sfrac{\bt^2}{\mpl^2} (a_1a_2a_3)^{-1} \left[ \sfrac{\langle\psi_y^2\rangle}{a_2^2} + \langle U(\psi)\rangle \right] \\[2mm]
    \sfrac{a''_3}{a_3} + \sfrac{a'_3}{a_3} \bigg( \sfrac{a'_1}{a_1} + \sfrac{a'_2}{a_2} \bigg) &= \sfrac{\bt^2}{\mpl^2} (a_1a_2a_3)^{-1} \left[ \sfrac{\langle\psi_z^2\rangle}{a_3^2} + \langle U(\psi)\rangle \right] \:.
\label{eq:scale_factors_numerical}
\end{align}
Finally, the equation of motion for the tensor perturbations is obtained as
\begin{align}
    v''_{ij} + K v_{ij} - \sum_{k=1}^3 \sfrac{\p_k^2 v_{ij}}{a_k^2} = (a_1 a_2 a_3)^{-1/2} \, \sfrac{\p_i \psi \, \p_j \psi}{a_i \, a_j}
\end{align}
where $K$ is defined in Eq.\ \autoref{eq:coeff_scales}.

These are the equations that will be solved numerically. The structure of the equations for the scalar field and the tensor perturbations are already in a suitable form for the leapfrog algorithm. However, the equations for the scale factors need a modification since the first time derivatives are one half step behind the corresponding variable at each step. In order to synchronize the variables and their first time derivatives we also keep their values from the previous step, meaning that we calculate the derivatives for this particular purpose as $a'_1(t) \approx [a'_1(t+\Delta t/2) + a_1(t-\Delta t/2)]/2$ where $\Delta t$ is the time step. We use those values to calculate the expression given in Eq.\ \autoref{eq:coeff_scales} as well.

\subsection{Initial Conditions}

In order to start the simulations we use the thin-wall approximation \cite{Coleman1977_1} to determine the initial profile of the scalar field, that is, we implement
\begin{equation}
    \psi(t=0, r) = \sfrac{1}{2} \bigg[ 1 - \tanh \! \bigg( \sfrac{r-R_{\rm{c}}M}{l_0 M} \bigg) \bigg]
\label{eq:thin-wall}
\end{equation}
where $R_{\rm{c}}$ and $l_0$ are the critical radius and the bubble wall length, respectively, which can be found from the following expressions \cite{Cutting2018}
\begin{equation}
    \psi(R_{\rm{c}}) = \sfrac{1}{2} \:, \qquad \psi(r^\pm) = \sfrac{1}{2} \bigg[ 1 - \tanh \! \bigg( \!\! \pm \sfrac{1}{2} \bigg) \bigg] \:, \qquad l_0 M = r^+ - r^- \:.
\end{equation}
Furthermore, the time derivative of the scalar field is taken initially to be zero, i.e. $\psi'(t=0)=0$. On the other hand, the bubble nucleation points in the lattice are randomly determined and the bubbles are nucleated simultaneously at the beginning of the simulations.

As for the scale factors we choose
\begin{align}
    a_1(t=0) = a_2(t=0) = a_3(t=0) = 1
\label{eq:sf_ini}
\end{align}
and in order to determine the initial values for the derivatives of the scale factors we use Eq.\ \autoref{eq:eins00} written with the new variables as
\begin{align}
    a'_1 a'_2 + a'_1 a'_3 + a'_2 a'_3 = \sfrac{\bt^2}{\mpl^2} \bigg[ \sfrac{1}{2} \! \left( \langle\psi_x^2\rangle \!+\! \langle\psi_y^2\rangle \!+\! \langle\psi_z^2\rangle \right) + \sfrac{1}{4} \big( a'_1 \!+\! a'_2 \!+\! a'_3 \big)^2 \langle \psi^2 \rangle + \langle U(\psi)\rangle \bigg]
\end{align}
with the help of Eq.\ \autoref{eq:sf_ini}. Moreover, assuming that $a'_1 = a'_2 = a'_3$ initially we get
\begin{align}
    a'_1(t=0) = a'_2(t=0) = a'_3(t=0) = \pm \sfrac{1}{3I_2-1} \sqrt{ \sfrac{1}{3} I_1 (1-3I_2)}
\end{align}
where
\begin{align}
    I_1 = \sfrac{\bt^2}{\mpl^2} \bigg[ \sfrac{1}{2} \Big( \langle\psi_x^2\rangle + \langle\psi_y^2\rangle + \langle\psi_z^2\rangle \Big) + \langle U(\psi)\rangle \bigg] \quad , \qquad I_2 = \sfrac{\bt^2}{\mpl^2} \langle \psi^2 \rangle \: .
\end{align}
Finally, we set $v_{ij}(t=0) = v'_{ij}(t=0) = 0$ for the tensor perturbations.

\section{DENSITIES, POWER SPECTRA, AND SHEAR SCALAR\label{sec:observables}}

Here we give the definitions for the densities, the power spectra for both the scalar field and the gravitational waves, and the shear scalar which will show the amount of anisotropy in simulations with different configurations. Starting with the densities for the scalar field we have
\begin{align}
    \bar{\rho}_{K} \equiv \sfrac{1}{2} \big\langle \dot{\phi}^2 \big\rangle \:, \qquad \bar{\rho}_{G} \equiv \sfrac{1}{2} \bigg\langle \sum^3_{k=1} \sfrac{\p_k^2\phi}{a_k^2} \bigg\rangle \:, \qquad \bar{\rho}_{V} \equiv \big\langle V(\phi) - V(\phi_{\rm{t}}) \big\rangle
\label{eq:densities_scalar_field}
\end{align}
which are the kinetic, the gradient, and the potential energies, respectively. On the other hand, for the energy density of the gravitational waves, the following expression is calculated 
\begin{equation}
    \bar{\rho}_{\rm gw}(\mf{x},t) = \sfrac{1}{8} \Mpl^2 \sum_{i,j} \Big\langle \dot{h}_{ij}(\mf{x},t) \, \dot{h}_{ij}(\mf{x},t) + \nb h_{ij}(\mf{x},t) \, \nb h_{ij}(\mf{x},t) \Big\rangle \:.
\label{eq:densities_gw}
\end{equation}
Although we keep the gradient terms explicitly, we need to emphasize that they almost have no effect on the results.

The power spectrum for the scalar field is expressed by
\begin{align}
    \P_\phi(\mf{k},t) = \sfrac{k^3}{2\pi^2} \big\langle \phi(\mf{k},t) \, \phi^*(\mf{k},t) \big\rangle
\label{eq:spectra_scalar}
\end{align}
and for the gravitational waves we use
\begin{align}
    \der{\Omega_{\rm{gw}}}{\ln k} = \sfrac{1}{3 H^2} \sfrac{k^3}{16\pi^2} \big( P_{\dot{h}}(\mf{k},t) + k^2 P_h(\mf{k},t) \big) \:.
\label{eq:spectra_gw}
\end{align}
We also implement the following normalization
\begin{align}
    \der{\Omega_{\rm{gw}}}{\ln k} \longrightarrow \sfrac{1}{(H_* R_* \Omega_{\rm{vac}})^2} \der{\Omega_{\rm{gw}}}{\ln k}
\end{align}
where $H_*$ is the average Hubble parameter value at the time of the transition, $R_*$ is the mean bubble separation equals to $(\V/N_{\rm{b}})^{1/3}$ in which $\V$ and $N_{\rm{b}}$ are the physical volume of the simulation box and the number of bubbles, respectively. 

We should emphasize that the above definitions for the power spectra are not entirely correct in the anisotropic case. However, we will represent them regardless and left those calculations to future studies. Because, in addition to investigating the mass scale for the scalar field, our main focus in this paper is to demonstrate the time evolution of the shear scalar defined as
\begin{align}
    \sg^2 = \sfrac{1}{6} \Big[ (H_1 - H_2)^2 + (H_1 - H_3)^2 + (H_2 - H_3)^2 \Big]
\label{eq:shear_scalar}
\end{align}
in order to quantify the anisotropy in the background. The Friedmann equation can then be written in the following form
\begin{align}
    H^2 = \sfrac{\sg^2}{3} + \sfrac{1}{3} \Mpl^{-2} \left[ \sfrac{1}{2} \left( \langle\dot{\phi}^2\rangle + \sum^3_{k=1} \sfrac{\langle (\p_i \phi)^2\rangle}{a_i^2} \right) + \langle V(\phi)\rangle \right]
\end{align}
or in terms of energy density parameters $1 = \Om_{\sg^2} + \Om_{\phi}$ where we have defined the energy density parameter for the shear scalar as follows
\begin{align}
    \Om_{\sg^2} \equiv \sfrac{\sg^2}{3 H^2}
\label{eq:shear_density}
\end{align}
and put all the terms of the scalar field into $\Om_{\phi}$ since we will not use it further.

\clearpage

\begin{table}[t]
\renewcommand{\arraystretch}{0.9}
\begin{tabular}{C{0.9cm}|C{1.4cm}|C{1.4cm}|C{1.4cm}|C{1.2cm}|C{1.2cm}|C{0.01cm}|C{1.2cm}|C{1.2cm}|C{1.4cm}|C{1.2cm}|C{1.2cm}}
\hline
\#& $N$                  & $N_{\rm{b}}$         & $\Mpl/M$             & $\lb$                & $\dt/M$                 & & $M_{\rm{t}}/M$ & $\phi_{\rm{t}}/M$ & $\rho_{\rm{vac}}/M^4$ & $R_{\rm{c}} M$ & $l_0 M$  \\ \hline\hline
1 & \multirow{6}{*}{640} & \multirow{3}{*}{320} & 1                    & \multirow{7}{*}{0.5} & \multirow{7}{*}{-1.632} & & \multirow{7}{*}{1.14} & \multirow{7}{*}{2.45} & \multirow{7}{*}{0.495} & \multirow{7}{*}{7.15} & \multirow{7}{*}{1.71} \\ \cline{1-1} \cline{4-4}
2 &                      &                      & 10                   &                      &                         & &  &   &   &   \\ \cline{1-1} \cline{4-4}
3 &                      &                      & \multirow{5}{*}{100} &                      &                         & &  &   &   &   \\ \cline{1-1} \cline{3-3} 
4 &                      & 10                   &                      &                      &                         & &  &   &   &   \\ \cline{1-1} \cline{3-3} 
5 &                      & 40                   &                      &                      &                         & &  &   &   &   \\ \cline{1-1} \cline{3-3} 
6 &                      & 600                  &                      &                      &                         & &  &   &   &   \\ \cline{1-2} \cline{3-3} 
7 & 1280                 & 320                  &                      &                      &                         & &  &   &   &   \\ \hline
\end{tabular}
\caption{Parameter values used to create different configurations for the simulations. $N$ and $N_{\rm{b}}$ are the number of grid points and the number of bubbles, respectively. From left to right, first five constants are the free parameters that have been chosen to test the dependency of the results and in accordance with the previous works of Refs.\ \cite{Cutting2018,Arapoglu2022}. The mass ($M_{\rm{t}}/M$) and the scalar field value ($\phi_{\rm{t}}/M$) in true vacuum, vacuum energy density ($\rho_{\rm{vac}}/M^4$), the critical radius ($R_c M$), and bubble wall thickness ($l_0 M$) are calculated as explained in the text. For $N=640$ ($N=1280$) we take $dx=0.44$ ($dx=0.22$).}
\label{tab:parameters}
\end{table}

\begin{figure*}[!b]
\centering

    \begin{tabular}{@{}c@{}}\hspace{-4mm}
	\includegraphics[trim={2.5cm 2.2cm 0 0},clip,width=.33\linewidth]{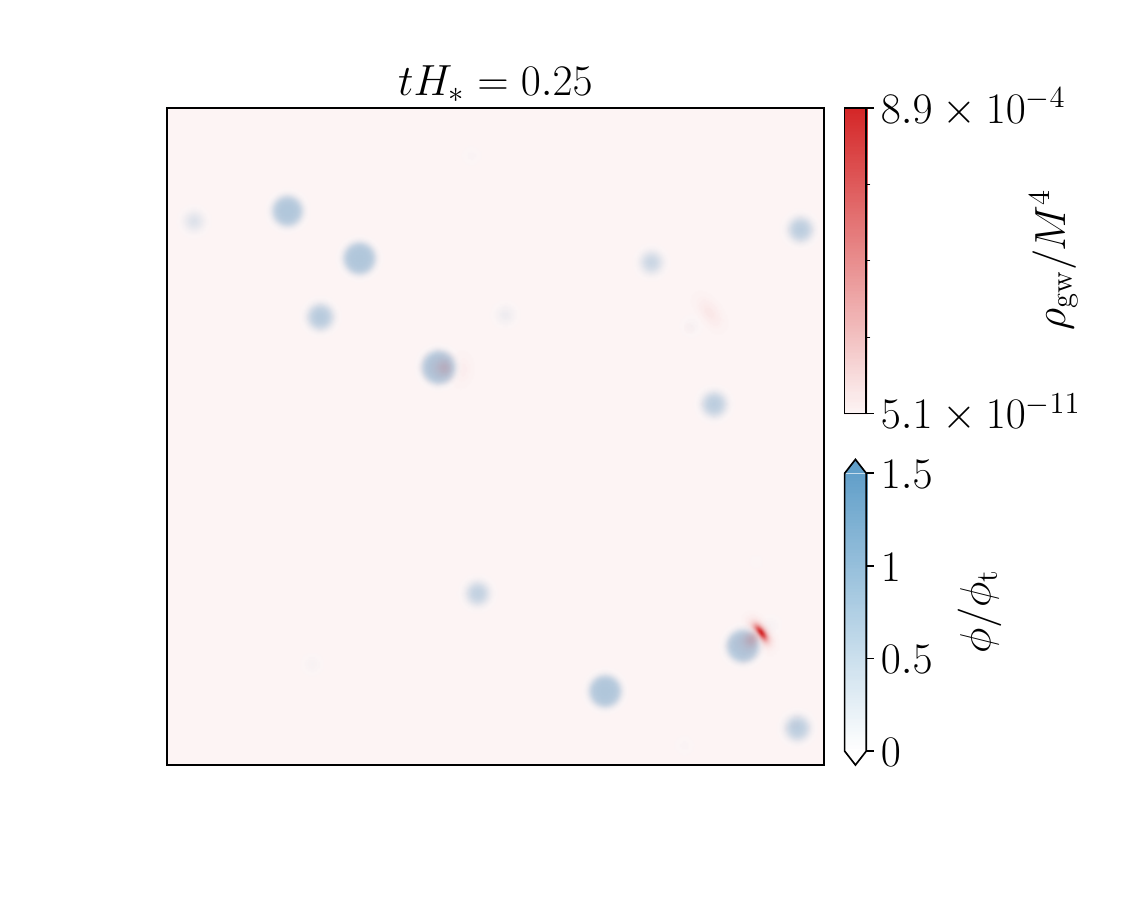}
    \end{tabular}
    \begin{tabular}{@{}c@{}}\hspace{-2mm}
	\includegraphics[trim={2.5cm 2.2cm 0 0},clip,width=.33\linewidth]{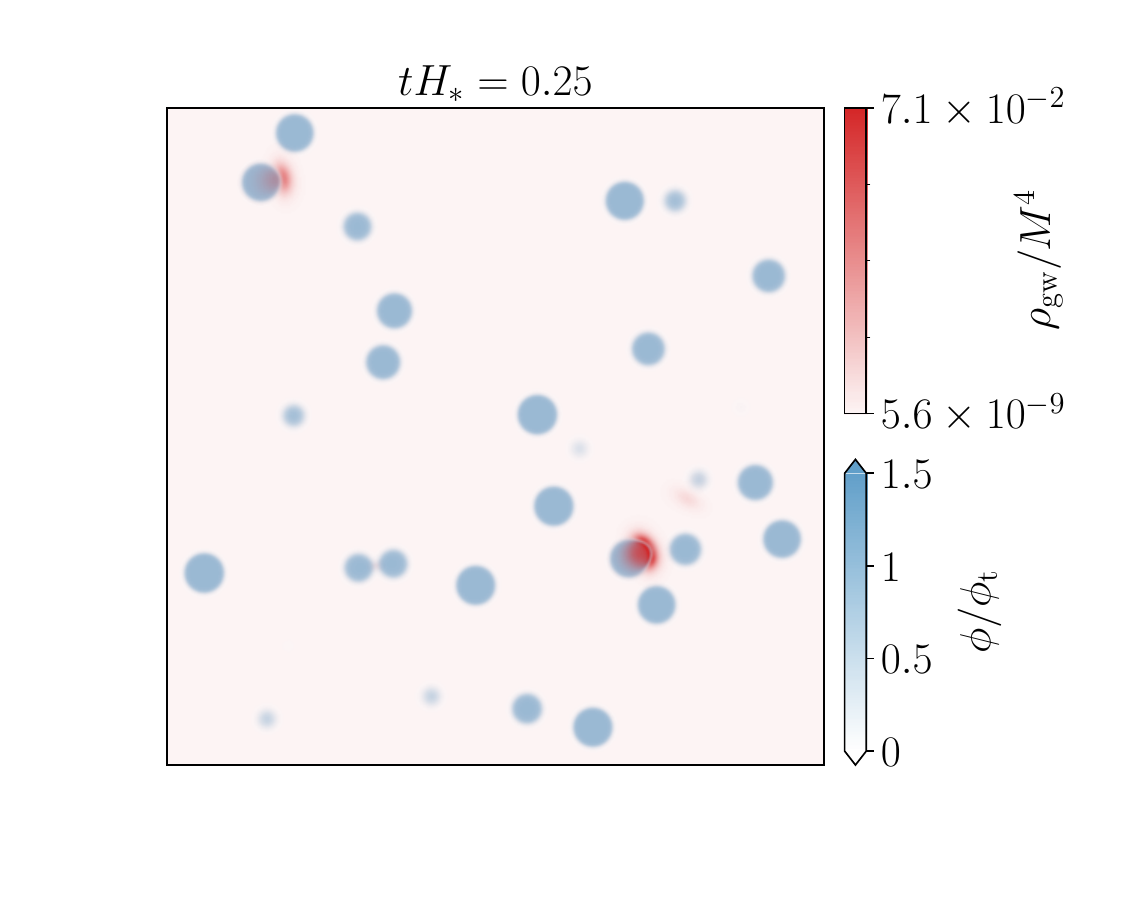}
    \end{tabular}
    \begin{tabular}{@{}c@{}}\hspace{-2mm}
	\includegraphics[trim={2.5cm 2.2cm 0 0},clip,width=.33\linewidth]{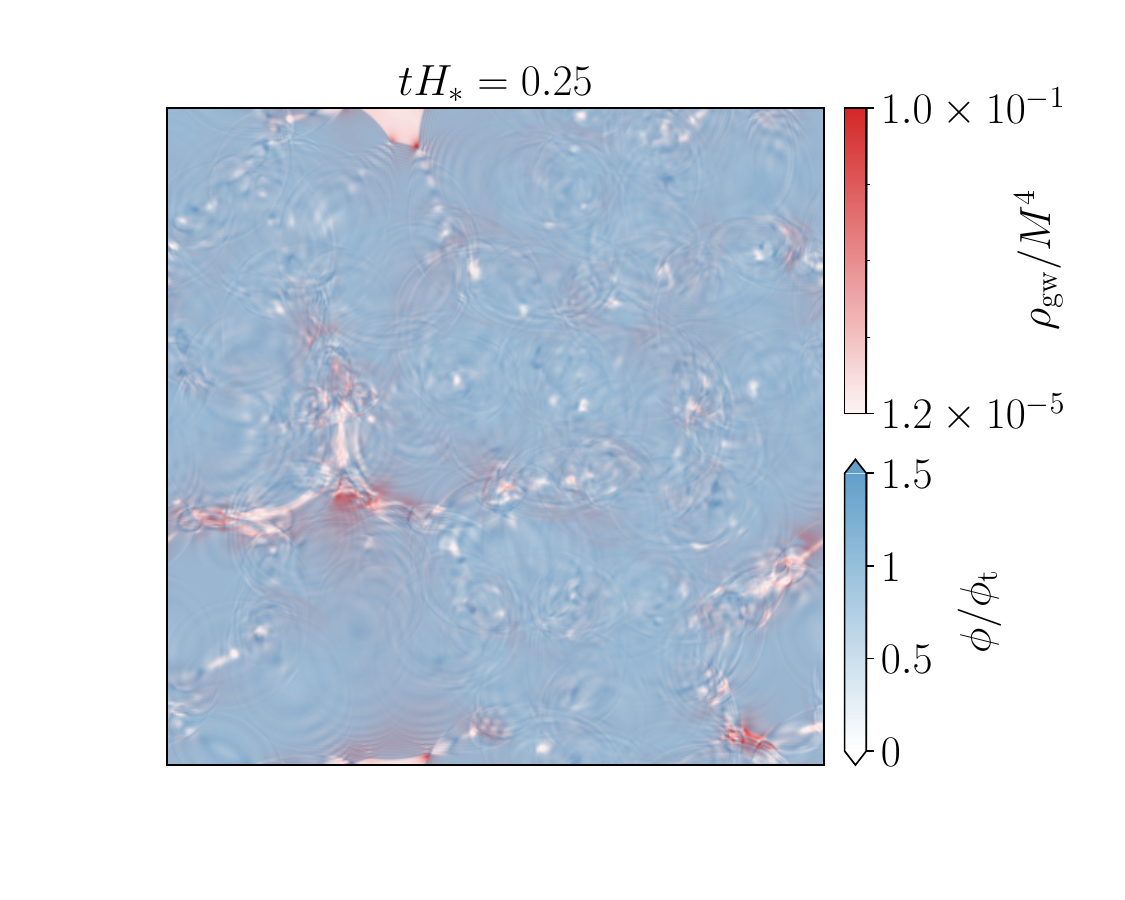}
    \end{tabular}

    \begin{tabular}{@{}c@{}}\hspace{-4mm}
	\includegraphics[trim={2.5cm 2.2cm 0 0},clip,width=.33\linewidth]{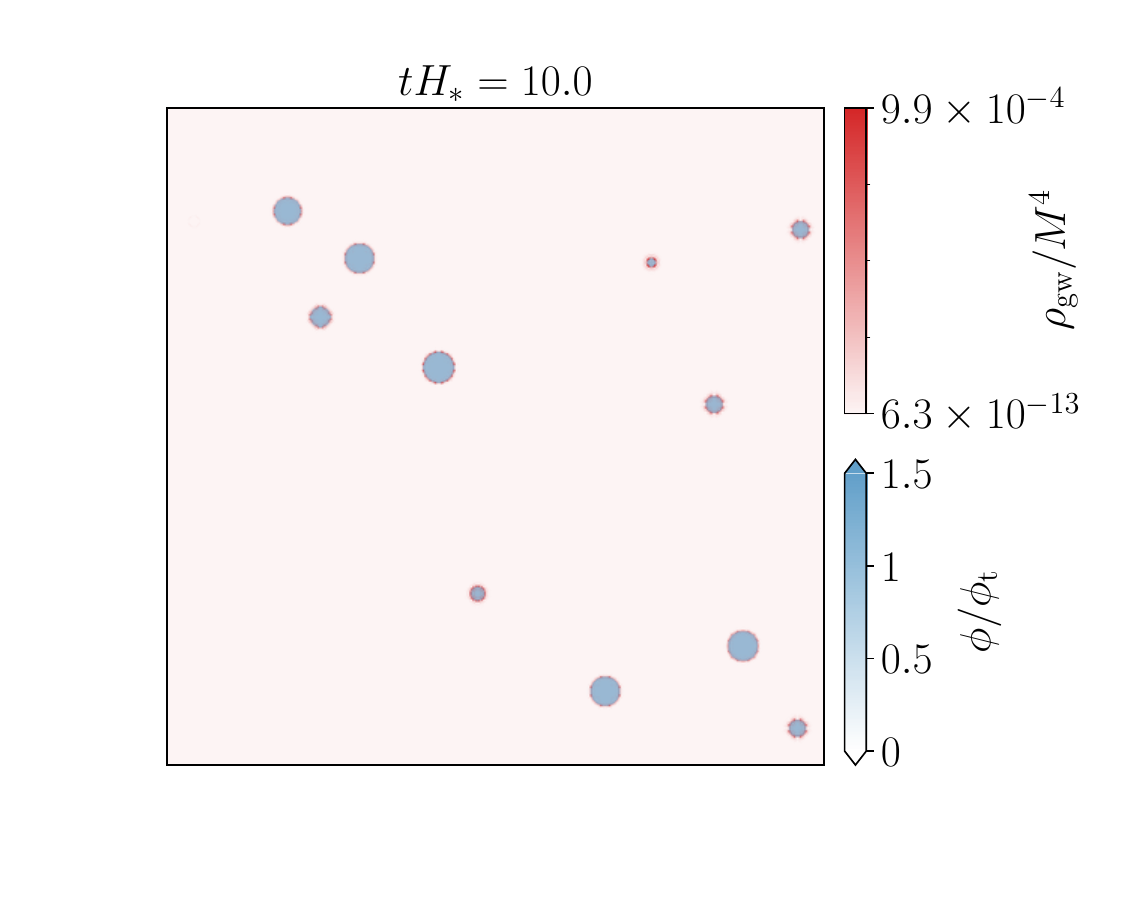}
    \end{tabular}
    \begin{tabular}{@{}c@{}}\hspace{-2mm}
	\includegraphics[trim={2.5cm 2.2cm 0 0},clip,width=.33\linewidth]{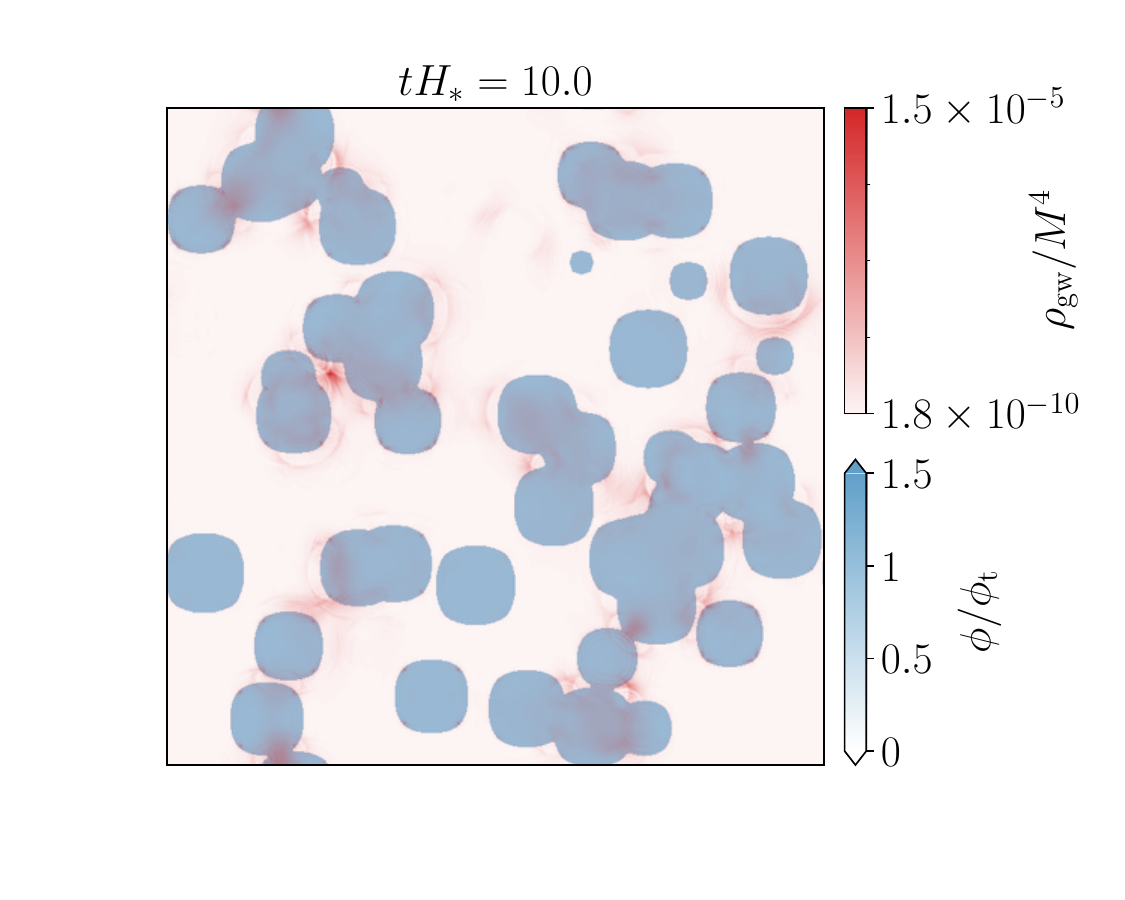}
    \end{tabular}
    \begin{tabular}{@{}c@{}}\hspace{-2mm}
	\includegraphics[trim={2.5cm 2.2cm 0 0},clip,width=.33\linewidth]{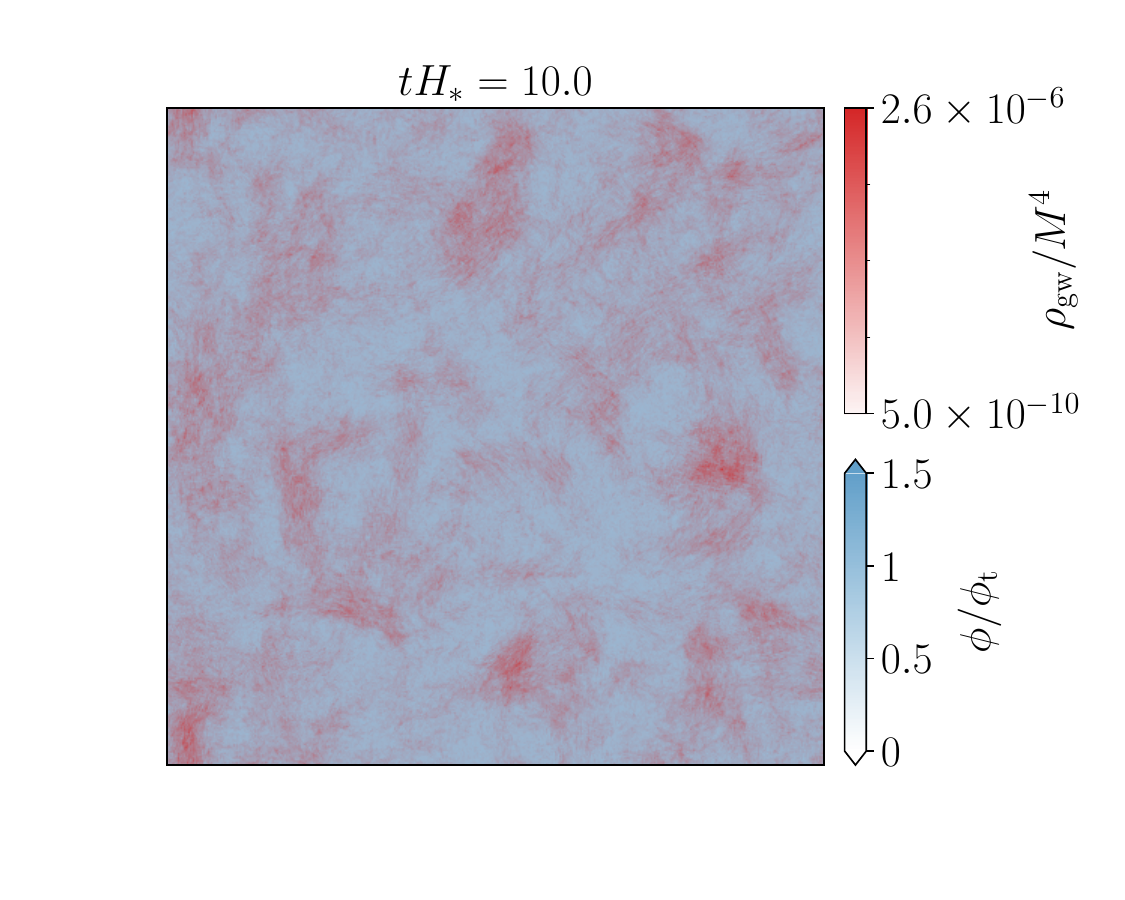}
    \end{tabular}

    \caption{Two-dimensional slices of the simulations with $\Mpl/M=1,10,100$ from left to right, respectively, for two different times. For all simulations, we take $N=640$ and $N_{\rm{b}}=320$. Note that the physical scales are different in each slice due to the difference in expansion rate.}
\label{fig:snapshots}
\end{figure*}

\clearpage

\section{SIMULATION RESULTS\label{sec:results}}

We have simulated several configurations of the model with different parameter values as listed in Table \autoref{tab:parameters}. In order to distinguish their effects on the results we have changed the number of nucleated bubbles (simulations 3-6) and the number of grid points (simulations 6, 7) as well as the mass scale (simulations 1, 2, 3) which is one of the main concern of this study in the context of free parameters. The values of the other constants that depend on the free parameters have also been calculated and given in the same table. The outcomes that will be reported and discussed in detail below are the Hubble parameters, densities, power spectra, and the shear scalar whose time evolution is another main interest of our work as it is one of the key indicators of anisotropy.

\begin{figure*}[!b]
\centering
        
    \begin{tabular}{@{}c@{}}\hspace{-2mm}
	\includegraphics[width=.49\linewidth]{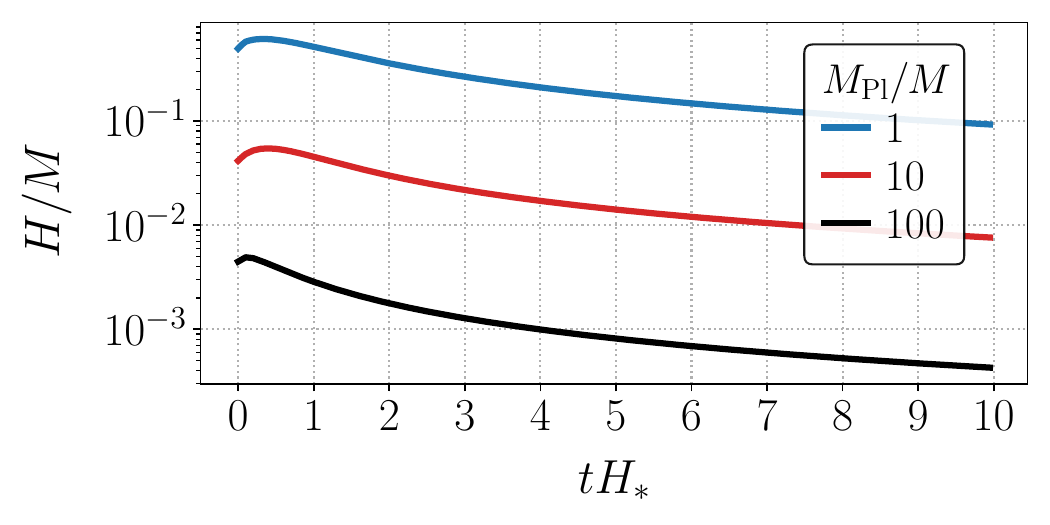}
    \end{tabular}
    \begin{tabular}{@{}c@{}}\hspace{1mm}
	\includegraphics[width=.49\linewidth]{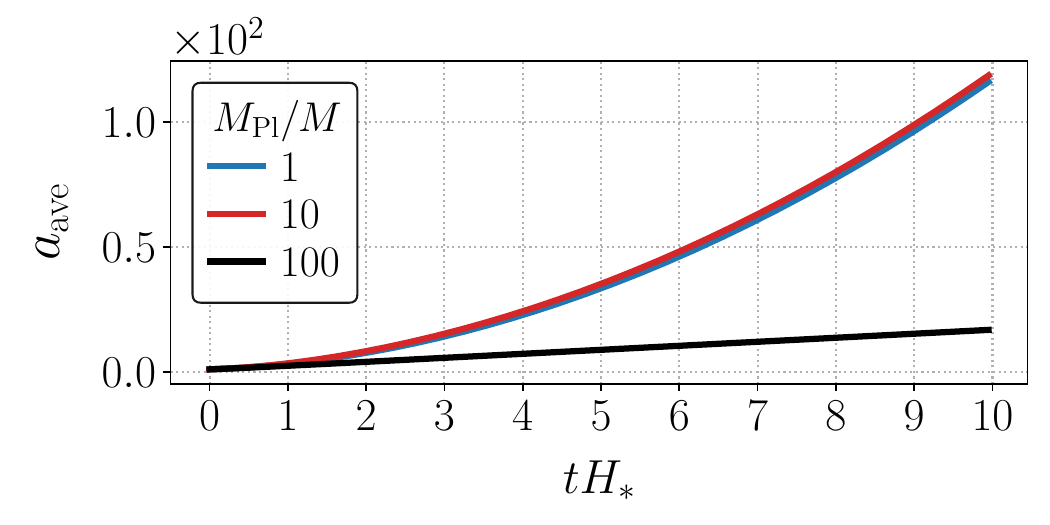}
    \end{tabular}

    \vspace{-3mm}
	
    \caption{Time evolution of the average Hubble parameters (left) and the scale factors (right) for the configurations with different mass scales. Here $N=640$, $N_{\rm{b}}=320$ for all simulations.}
\label{fig:hubbles}
\end{figure*}

First of all, we have run simulations to see the mass scale at which the phase transition can be completed successfully. We see from two-dimensional slices of the simulations represented in Fig.\ \autoref{fig:snapshots} that the phase transition does not complete for the configurations with $\Mpl/M=1,10$. This is due to the fact that the expansion rate is higher for relatively small values of $\Mpl/M$ as one can deduce this result from the right-hand side of Eqs.\ \autoref{eq:scale_factors_numerical}. The high expansion rate causes that universe to outgrow with a speed much more than the enlargement of the bubbles, therefore, the bubble collision phase either can not be completed entirely or does not happen at all. For $\Mpl/M=10$ the bubbles expand for a while at the start of the simulation and the ones close to each other collide partially, but then, the expansion rate of the universe eventually dominates the dynamics, whereas the bubble collision does not even occur in the case of $\Mpl/M=1$. On the other hand, for $\Mpl/M=100$ we notice that this kind of effect does not take place and the bubble collision phase is completed successfully in a time less than $H_*^{-1}$. At this point it is necessary to emphasize that this mass value is beyond even the GUT scale let alone the EW phase transition epoch, which are around $10^{15}$ GeV and $100$ GeV, respectively. Therefore, the completion of a phase transition at either GUT or EW scale is not affected by the expansion of the Universe. On the other hand, this inference can also be supported by the time evolution of the average scale factors and the average Hubble parameters represented in Fig.\ \autoref{fig:hubbles}. Higher mass ratios correspond to lower expansion rates in comparison at the same time scale. However, we should note that the scale factor of the case with $\Mpl/M=10$ is less than that of $\Mpl/M=1$ at the initial stages of their time evolution although it becomes slightly larger afterwards as can be seen in the figure as well. Nevertheless, the time evolution of those two curves are very similar and they differ from that of $\Mpl/M=100$ almost two orders of magnitude. However, the cases with $\Mpl/M=1,10$ are the scenarios we did not take into account further as they do not fulfill the requirements to complete the phase transition.

\begin{figure*}[!t]
\centering

    \begin{tabular}{@{}c@{}}\hspace{-6mm}
	\includegraphics[width=.49\linewidth]{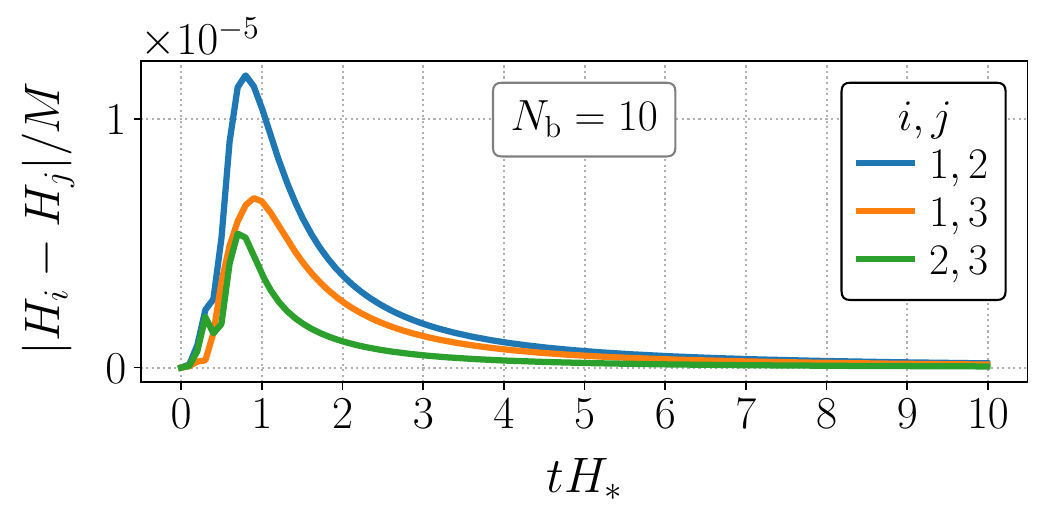}
    \end{tabular}
    \begin{tabular}{@{}c@{}}\hspace{3mm}
	\includegraphics[width=.49\linewidth]{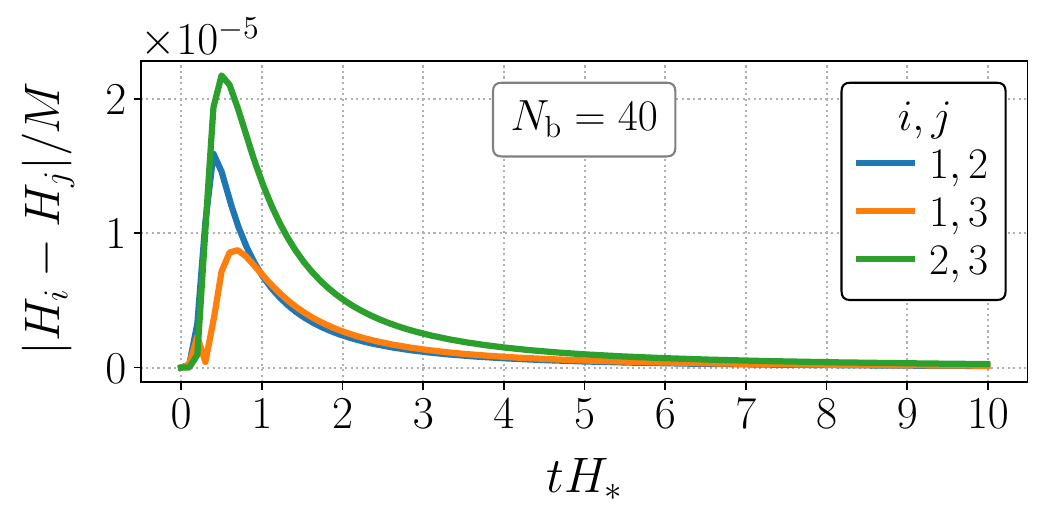}
    \end{tabular}

    \vspace{-3mm}
    
    \begin{tabular}{@{}c@{}}\hspace{-6mm}
	\includegraphics[width=.49\linewidth]{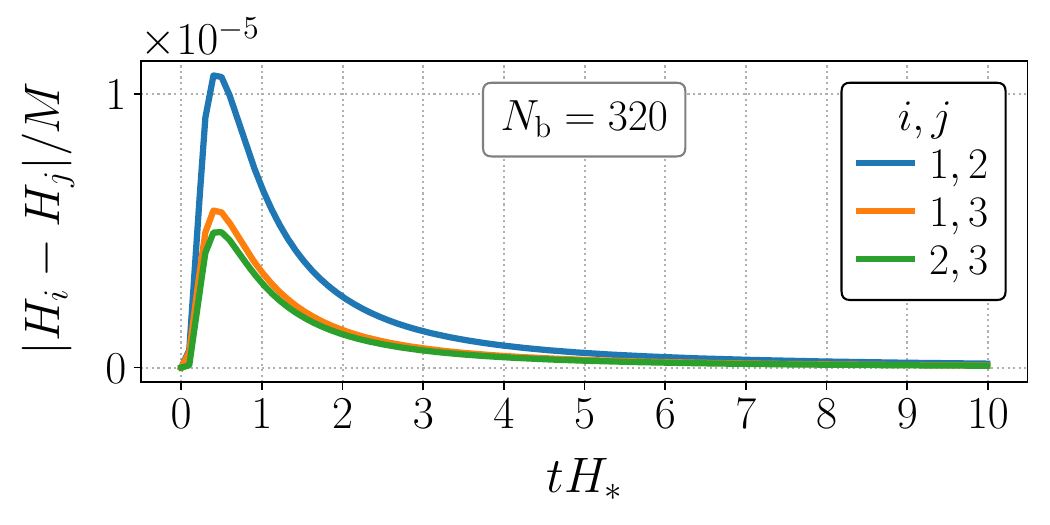}
    \end{tabular}
    \begin{tabular}{@{}c@{}}\hspace{3mm}
	\includegraphics[width=.49\linewidth]{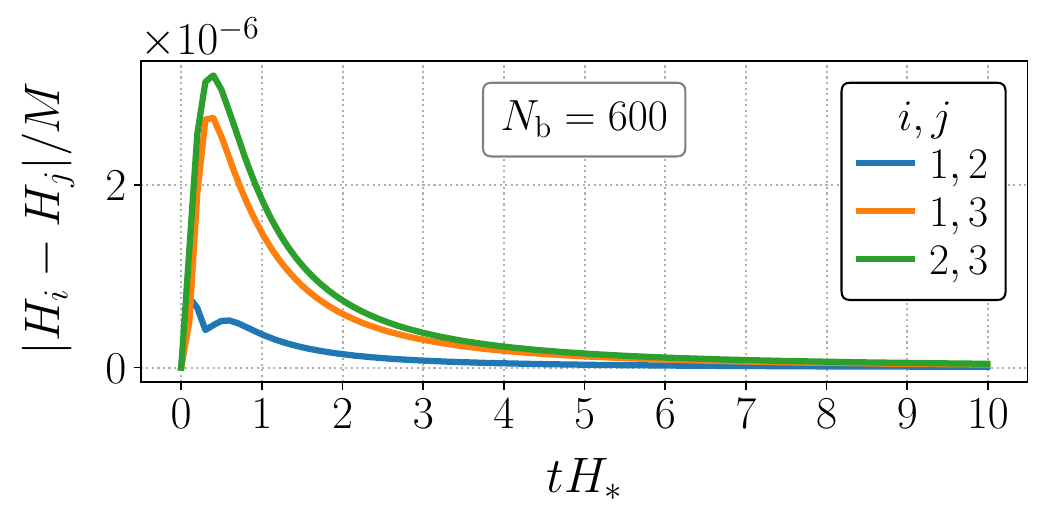}
    \end{tabular}

    \vspace{-3mm}

    \caption{Time evolution of the absolute differences between the directional Hubble parameters for the configurations with different values of number of bubbles as depicted in the figures. For all simulations we take $N=640$ and $\Mpl/M=100$.}
\label{fig:dir_hubbles_Nb}
\end{figure*}

\begin{figure*}[!t]
\centering

    \begin{tabular}{@{}c@{}}\hspace{-6mm}
	\includegraphics[width=.49\linewidth]{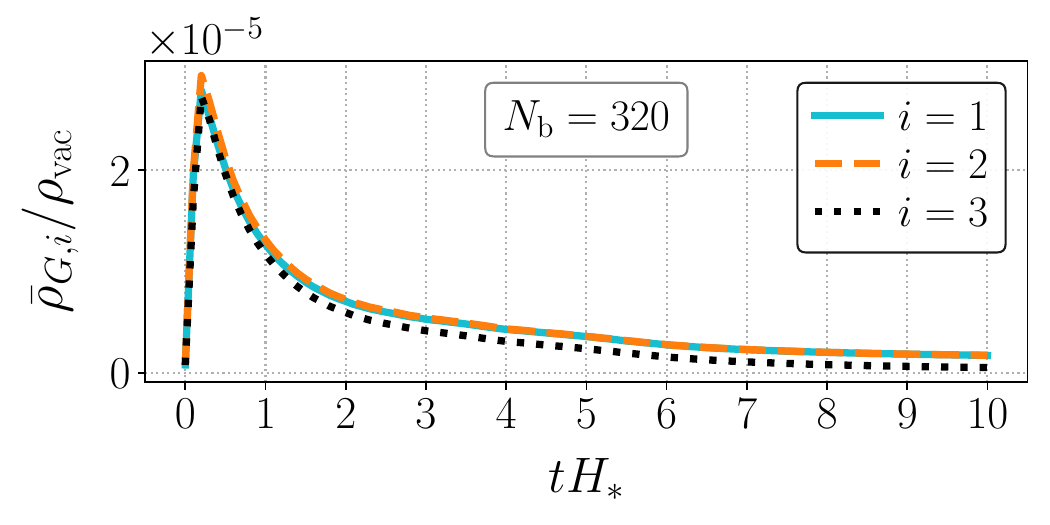}
    \end{tabular}
    \begin{tabular}{@{}c@{}}\hspace{3mm}
	\includegraphics[width=.49\linewidth]{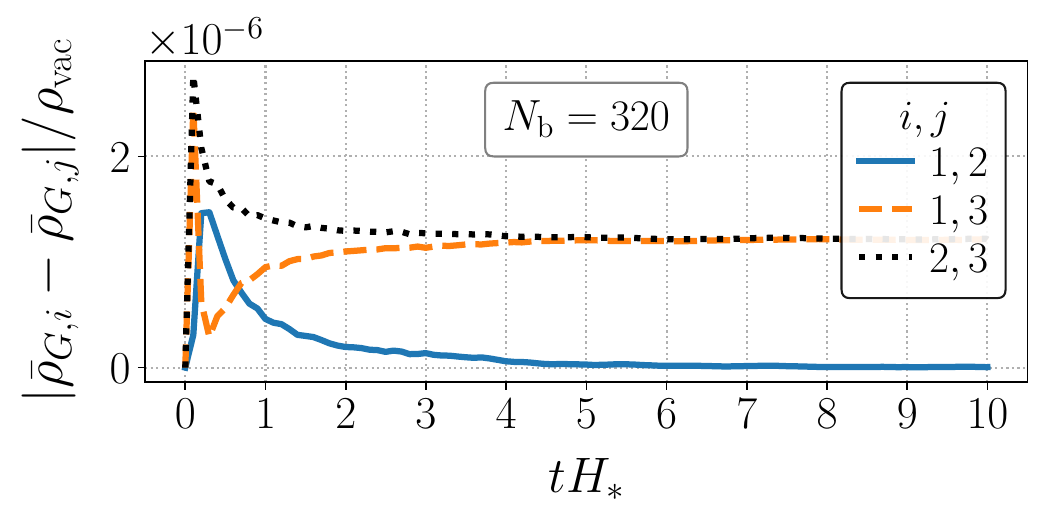}
    \end{tabular}

    \vspace{-3mm}
		
\caption{Time evolution of in the directional average gradient energies (left) and the absolute differences (right) for a sample configuration with $N_{\rm{b}}=320$, $N=640$, and $\Mpl/M=100$.}\vspace{3mm}
\label{fig:dir_densities_Nb}
\end{figure*}

\begin{figure*}[!b]
\centering
    \vspace{5mm}
    \begin{tabular}{@{}c@{}}\hspace{-4mm}
	\includegraphics[width=.49\linewidth]{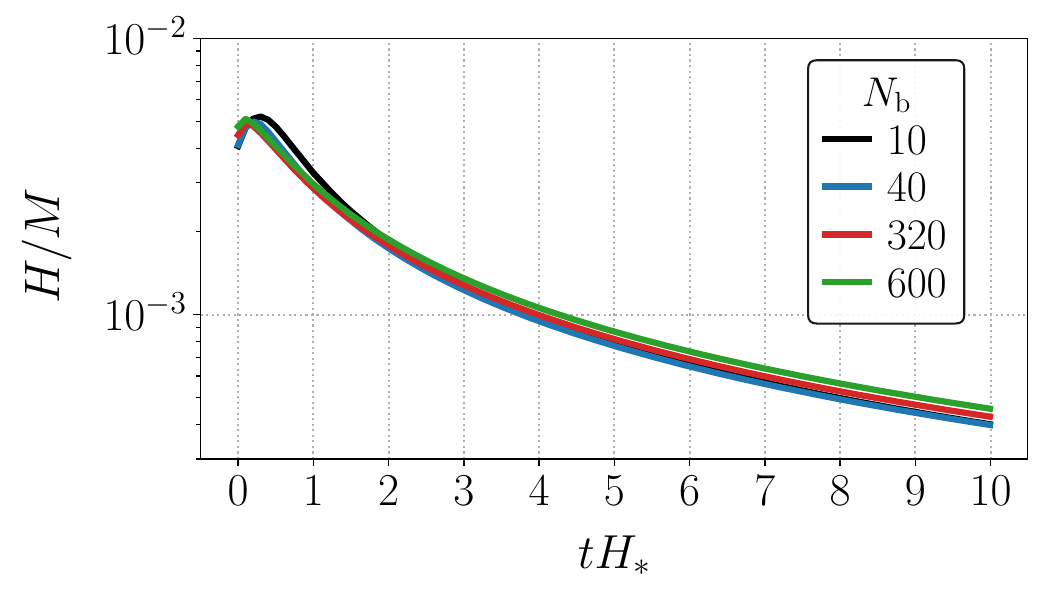}
    \end{tabular}
    \begin{tabular}{@{}c@{}}\hspace{3mm}
	\includegraphics[width=.49\linewidth]{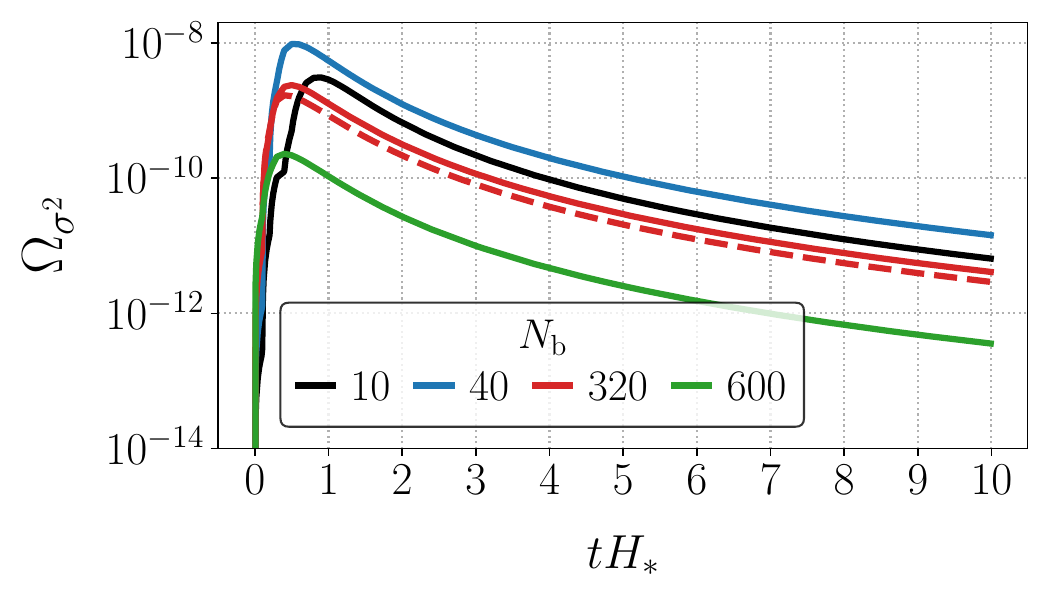}
    \end{tabular}

    \vspace{-3mm}

    \caption{Time evolution of the Hubble parameters (left) and the shear scalar (right) for the simulations with different values of number of bubbles and grid points. Here we take $\Mpl/M=100$ for all simulations. For the solid lines, $N=640$, and $dx=0.44$ whereas $N=1280$, $dx=0.22$, and $N_{\rm{b}}=320$ for the dashed line.}
\label{fig:hubble_shear_Nb}
\end{figure*}

Since we have confirmed that it is required to choose $\Mpl/M \gtrsim 100$ roughly in order the transition to complete, now we will investigate the effect of the other parameters on the outcomes such as the number of initiated bubbles, $N_{\rm{b}}$, and the number of grid points, $N$. We examine the impact of $N_{\rm{b}}$ through four different simulations with $N_{\rm{b}}=10,40,320,600$ fixing $\Mpl/M=100$ and $N=640$ for all runs. The results are shown in Fig.\ \autoref{fig:dir_hubbles_Nb} for the differences in the directional Hubble parameters, in Fig.\ \autoref{fig:dir_densities_Nb} for a sample of the gradient energy densities together with the differences in the directional components, and in Fig. \autoref{fig:hubble_shear_Nb} for the average Hubble parameters and the energy density parameter of the shear scalar. As seen from Fig.\ \autoref{fig:dir_hubbles_Nb} although there is only one order of magnitude between them, the maximum value in the differences decreases with increasing number of bubbles except for the case of $N_{\rm{b}}=10$ for which we have found that the transition does not complete and gives results similar to the case with $\Mpl/M=10$ given in Fig.\ \autoref{fig:snapshots}. Regarding the difference in the gradient energies, we have found that they are in the same order of magnitude for all four different simulations and we have represented one example of them in Fig.\ \autoref{fig:dir_densities_Nb}. As seen from the figure, the directional quantities peak at early stages of the simulations corresponding to the bubble collision phase and then decreases smoothly throughout the run.

\begin{figure*}[!t]
\centering

    \begin{tabular}{@{}c@{}}\hspace{-4mm}
	\includegraphics[width=.55\linewidth]{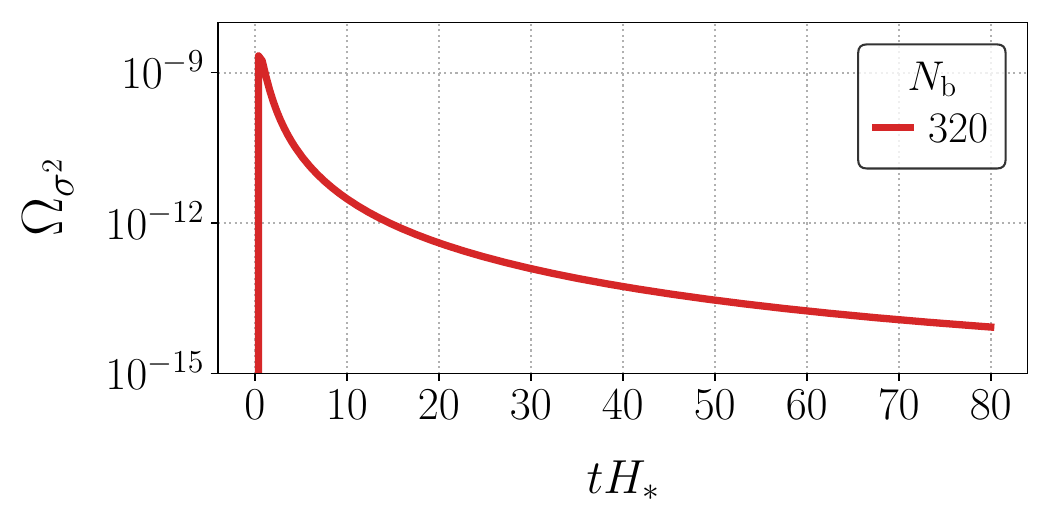}
    \end{tabular}

    \vspace{-3mm}

    \caption{Time evolution of the shear scalar of a simulation with $\Mpl/M=100$, $N=640$, and $N_{\rm{b}}=320$ for a longer run in comparison with the ones given in Fig.\ \autoref{fig:hubble_shear_Nb}.}
\label{fig:shear_Nb_long_run}
\end{figure*}

Together with the corresponding Hubble parameters the results for the shear scalar defined in Eq.\ \autoref{eq:shear_scalar} are represented in Fig.\ \autoref{fig:hubble_shear_Nb} in terms of its energy density parameter given in Eq.\ \autoref{eq:shear_density}. The curves show that the value of the shear scalar increases to some extent with decreasing number of bubbles and then starts to get smaller after some value in accordance with discussion about the difference between the components of the directional Hubble parameters in the previous paragraph and we should recall that for $N_{\rm{b}}=10$ the transition is not accomplished. The shear scalar has almost the same shape throughout its time evolution in different simulations as if it was shifted depending on the number of bubbles. Nevertheless, the maximum values occur around $10^{-8}-10^{-10}$ right after the completion of the bubble collision phase and then within our time scale for the simulations it reaches $10^{-11}-10^{-12}$ decreasing gradually. Moreover, we have also provided a result drawn with a dashed line on the right panel of Fig.\ \autoref{fig:hubble_shear_Nb} in order to check the effect of resolution of the simulation box. We see that the shear scalar gets slightly smaller for higher resolution with the same number of bubbles. Nevertheless, for all cases the shear scalar increases during the bubble collision phase and then it decreases as the scalar field oscillates around its true vacuum. Since we do not expect an anisotropic structure to develop in this configuration at late times, we can conclude that the maximum value for the shear scalar energy density parameter that we have found is around $10^{-8}$. Additionally, we have also provided the result of a longer run for a simulation with $N_{\rm{b}}=320$, $N=640$, and $\Mpl/M=100$ in Fig.\ \autoref{fig:shear_Nb_long_run}. We see that the energy density parameter of the shear scalar reaches values around $10^{-14}$ at the end of that simulation. We need to note that this value already matches one of the most stringent constraint on today's value for the energy density parameter of the shear scalar \cite{Akarsu2019}, that is, in the order of $10^{-15}$, and, moreover, it continues to decrease.

\begin{figure*}[!h]
\centering

    \begin{tabular}{@{}c@{}}\hspace{-4mm}
	\includegraphics[width=.49\linewidth]{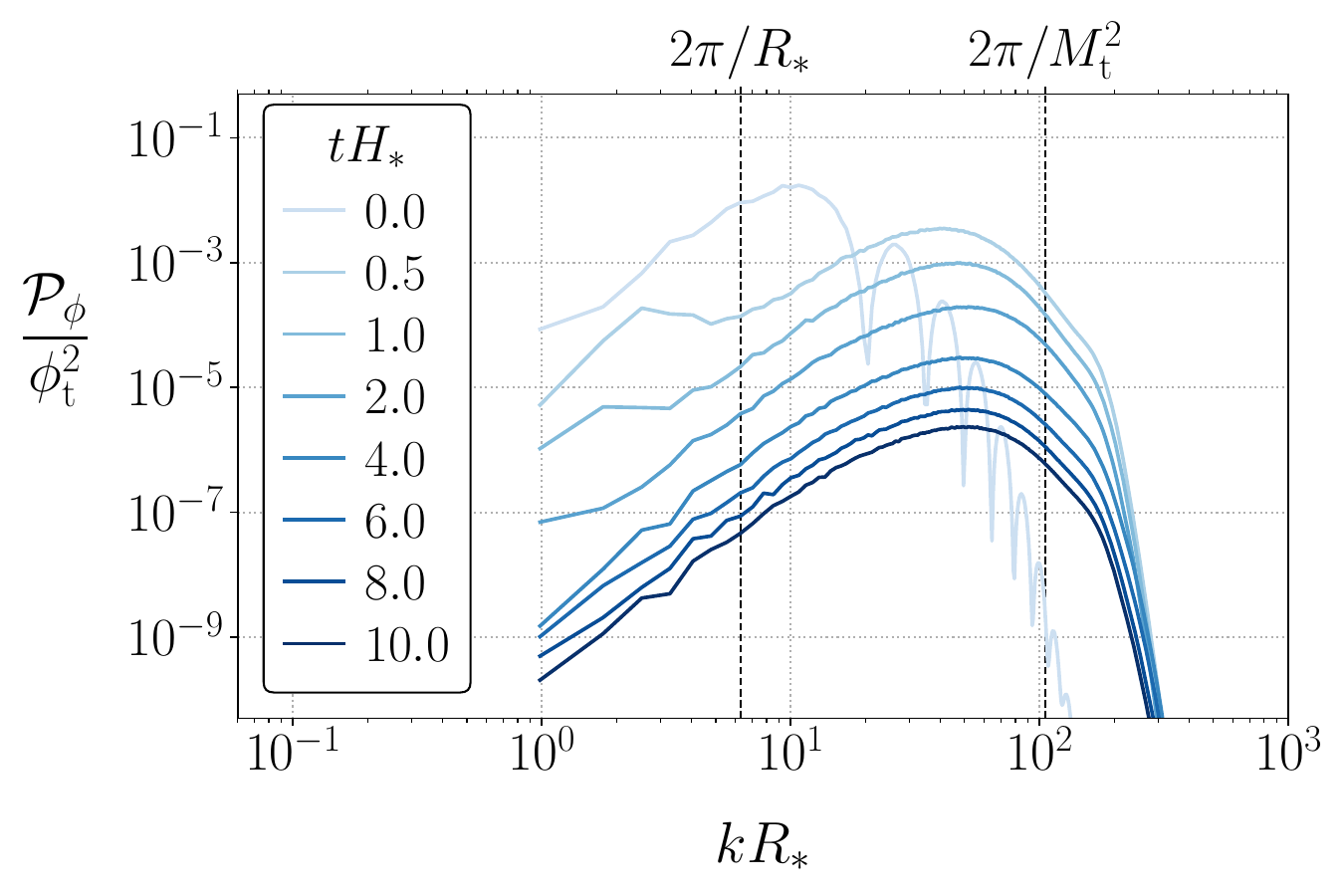}
    \end{tabular}
    \begin{tabular}{@{}c@{}}\hspace{1mm}
	\includegraphics[width=.49\linewidth]{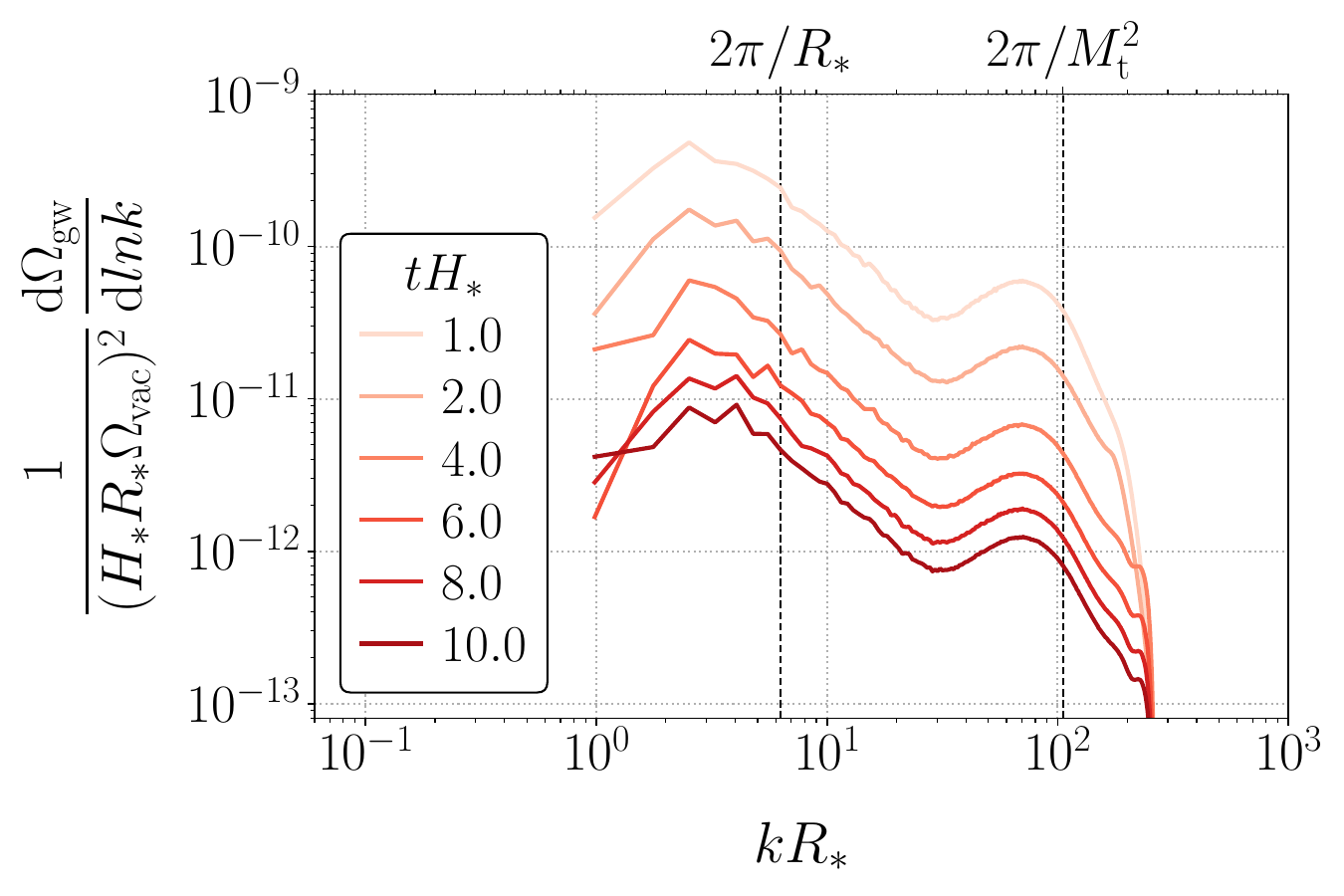}
    \end{tabular}

    \vspace{-3mm}

    \caption{Power spectrum of the scalar field (left) and the GW energy density (right) for a configuration with $N_{\rm{b}}=320$. Here we take $N=640$ and $\Mpl/M=100$.}
\label{fig:spectrum}
\end{figure*}

The results for the power spectra of the scalar field and the GW energy density, defined in Eqs.\ \autoref{eq:spectra_scalar} and \autoref{eq:spectra_gw} respectively, are represented in Fig.\ \autoref{fig:spectrum}. We have shown only one example for the case of $N_{\rm{b}}=320$ due to the fact that change in number of bubbles does not effect the shape of the spectrum neither for the scalar field nor for the GW energy density. We understand from the figures that the bubble collision phase is completed successfully before $t H_* = 1$ since the scalar field already oscillates around its true vacuum corresponding to a peak of its power spectrum near the mass value $M_{\rm{t}}$ and the power spectrum of the GW energy density develops secondary peak there as well. The overall magnitude in both power spectra decreases due to the expansion while keeping the same shape. Therefore, the characteristic shapes for both power spectra are the same with the results of previous works \cite{Cutting2018,Arapoglu2022}. However, as we have mentioned before the tensor perturbations should be investigated in detail for Bianchi Type-I model and, in accordance with the spirit of the model, possible anisotropies in the power spectra with compatible definitions are needed to take into consideration which we left for future studies.

\section{CONCLUSION\label{sec:conclusion}}

In this paper, we have examined the cosmological first-order vacuum phase transitions in an anisotropic expanding universe modeled by Bianchi Type-I metric. To do this we have used a model with a scalar field that is minimally coupled to the gravity and has a typical potential for the first-order phase transitions. After representing the main equations in their analytical forms, we have put them into numerical set in accordance with the leapfrog algorithm. Then, we have integrated the equations of motion for the scalar field and for the directional scale factors as well as for the tensor perturbations, the results of which are valid up to a gauge transformation due to the fact that in Bianchi Type-I model the TT gauge should be modified \cite{Miedema1993,Cho1995}. In addition to that it is also important to check the anisotropy in the GW power spectrum to either validate or eliminate a model or the source of the signal through possible upcoming observations even by taking the periodicity of the simulation box into account \cite{Racz2020}. Nevertheless, main purpose of this work was to find out the mass scale at which the bubble collision phase is accomplished and, additionally, to track the anisotropy by determining the behavior of the shear scalar defined in Eq.\ \autoref{eq:shear_scalar}, in other words, to consider the anisotropy in the background evolution due to the scalar field responsible from the transition.

We have run several simulations with different number of initiated bubbles which determines the initial conditions and correspondingly has the major impact for the time evolution of all variables. In addition to that due to the computational costs we have simulated only one configuration with higher resolution and the one with a longer run in comparison with the others. Before investigating the shear scalar, we have represented the results for three simulations with different mass scales, namely $\Mpl/M=1,10,100$, which have shown that the phase transition does not complete for the runs roughly $\Mpl/M \lesssim 100$. In those cases either the bubbles expand for a while and then the expansion of the universe prevents them to coalesce entirely or they do not find a chance to collide at all because of the expansion of the universe. We have given the results of examples for those two cases with the mass scales of $\Mpl/M=10$ and $\Mpl/M=1$, respectively, in Fig.\ \autoref{fig:snapshots} together with the case of $\Mpl/M=100$ that was adopted for the rest of the simulations. We did not use mass scales greater than that because of the computational costs and, moreover, this value is enough to examine the anisotropy in first place due to the fact that higher rates for $\Mpl/M$ suppress the expansion of the Universe more and more. 

After determining the order of minimum mass scale, that is $\Mpl/M \approx 100$, at which the phase transition can be completed successfully, we have run simulations to determine the time evolution of the energy density parameter for the shear scalar by examining the effect of different initial conditions created through different number of initiated bubbles. But before this we have shown in Fig.\ \autoref{fig:dir_hubbles_Nb} that absolute differences in the directional Hubble parameters are in the order of $10^{-5}$ for $N_{\rm{b}}=10, 40, 320$ while it is around $10^{-6}$ for $N_{\rm{b}}=600$ at most. Additionally, we have also provided the directional gradient energies and their differences in Fig.\ \autoref{fig:dir_densities_Nb} for the same configurations with the number of bubbles mentioned and have shown that their differences are in the order of $10^{-6}$.

In Fig.\ \autoref{fig:hubble_shear_Nb} we have presented the results for the Hubble parameters and the shear scalars. Moreover, we have also given the outcomes for a longer run of a specific configuration in Fig.\ \autoref{fig:shear_Nb_long_run}. As indicated before from the results of the directional Hubble parameters it seems that the relatively small number of bubbles give rise to high values for the shear scalar except for the case of $N_{\rm{b}}=10$ which seems to be a counter example for this conclusion at first glance, but the bubble collision phase is not completed for that simulation. Therefore, as one may guess before, in addition to the mass scale, the proportion of the number of initiated bubbles to whole simulation box is another quantity that also determines whether a phase transition can be completed or not. This can also be seen from Fig.\ \autoref{fig:hubble_shear_Nb} for the Hubble parameters where the case of $N_{\rm{b}}=10$ is different from the others at the beginning of the simulation. Nevertheless, we have found that before decreasing smoothly, the energy density parameter for the shear scalar gains a peak between $10^{-8}-10^{-10}$ which occurs at bubble collision phase. With the aforementioned longer run we have shown that $\Om_{\sg^2}$ becomes close to one of the constraints obtained for its today's value \cite{Akarsu2019}. Additionally, it seems that the expansion of the Universe does not effect the phase transition for a typical mass scales of $\Mpl/M \gtrsim 100$ with a fairly distributed number of initiated bubbles, since hereby we have tested impact of the expansion itself as well besides the anisotropy.

\section*{Acknowledgements}

This work is supported by The Scientific and Technological Research Council of Türkiye (TÜBİTAK) through grant number 121F066. Computing resources used in this work were provided by the National Center for High Performance Computing of Türkiye (UHeM) under grant number 5013072022 and the simulations were partially performed at TUBITAK ULAKBIM, High Performance and Grid Computing Center (TRUBA resources).

\clearpage

\bibliographystyle{apsrev4-2}
\bibliography{references}

\end{document}